\documentclass[12pt]{article}
\usepackage{amsmath}
\begin{document}
\begin{flushright}
USITP-98-07\\
June 1998
\end{flushright}
\bigskip
\Large
\begin{center}
{\bf Generating new supergravity solutions} \\ 
{\bf using Ehlers-Harrison-type transformations}\\
\bigskip

\normalsize
\bigskip
\bigskip

Henrik Gustafsson\footnote{e-mail: henrik@physto.se} 
and Parviz Haggi-Mani\footnote{e-mail: parviz@physto.se}\\
\smallskip
{\it ITP\\
University of Stockholm\\
Box 6730, Vanadisv\"agen 9\\
S-113 85 Stockholm\\
SWEDEN}\\
\end{center}
\vspace{5cm}
\normalsize

{\bf Abstract:} 
The technique of generating new solutions to $4D$ 
gravity/matter systems
by dimensional reduction to a $\sigma$-model is 
extended to supersymmetric
configurations of supergravity. The conditions required 
for the preservation of supersymmetry under isometry 
transformations in the $\sigma$-model 
target space are found. Some examples 
illustrating the technique are
given. 
\bigskip\bigskip\bigskip
\thispagestyle{empty}\eject
\setcounter{page}{1}

\section{Introduction}

Methods   of generating  new solutions   
from known old ones in gravity/matter systems 
have been discussed extensively and the techniques 
employed go back to Ehlers \cite{Ehl57}, Harrison \cite{Har68}, 
Neugebauer, Kramer \cite{Kra69}, Kinnersley \cite{Kin73}, 
and Geroch \cite{Ger71}. The cases 
of axion/dilaton and dilaton systems with electromagnetic 
fields present were discussed rather recently by \cite{Gal94} 
and \cite{Jen95}, respectively. 
In this approach one considers, {\it e.g.}, $4D$  space-times with 
a non-null Killing vector field.
The  solutions are 
described by $3D$ gravity coupled to a nonlinear $ \sigma$-model. 
The target space of the $ \sigma$-model admits isometries 
which are then used to generate new solutions.      

In this paper we discuss in detail  
supergravity  systems
and investigate the  possibility of generating new 
supersymmetric solutions.
In this case, the equations of motion can be 
found from a $3D$ supersymmetric $\sigma$-model coupled 
to supergravity. This is obtained by dimensional 
reduction of $4D$ supergravity.
The reduction 
here is a simple or na\"{\i}ve one, based on the 
assumption that all the fields are independent of
one coordinate which can be  chosen  to be along a space- or 
a time-like direction. For a supersymmetric 
$\sigma$-model coupled to gravity to exist in $3D$
the target space 
is restricted to be either (locally) K\"a{}hler or quaternionic
depending on the number of supersymmetries.
The isometries of the $\sigma$-model are then used  
 to generate new supersymmetric solutions from a given
supersymmetric ``seed'' solution.
 In order to guarantee the supersymmetry of
the new solutions, the Killing spinor(s) of the 
seed solution must be
independent of the isometry direction used to reduce
the original Lagrangian. In the context of superstring theory 
the same condition has been found
necessary for T-duality to preserve supersymmetry \cite{Bak,Kall}.
The dimensionally reduced Killing spinors are then invariant under 
the target space isometry transformations of the $\sigma$-model.

This paper is organised as follows: In section 2 we 
describe the problem in general. In section 3 we explicitly  
reduce the $N=1,\;\;D=4$ supergravity action to  $N=2,\;\;D=3$
supergravity coupled to a supersymmetric nonlinear $\sigma$-model, 
find the isometries, and give examples.
In section 4 we reduce the bosonic sector  
of  $N=2,\;\;D=4$ supergravity which  results in a $\sigma$-model 
with a quaternionic target space structure. It can therefore be 
extended to a supersymmetric $\sigma$-model. We also  generate
some new solutions to illustrate the method. Our conclusions are
summarized in section 5.

\section{ Generalities }

In this section we describe  the  technique of generating 
solutions
via dimensional reduction to a $\sigma$-model 
\cite{Ehl57}-\cite{Jen95},
 and extend it to the supersymmetric case.

The action\footnote{
Throughout this paper we use a $4D$ metric with signature 
$(-,+,+,+)$.
Curved indices  are denoted by $M,N,\ldots$ and tangent 
space indices
by $A,B,\ldots$. In $3D$ these are denoted $\mu,\nu,\ldots$ 
and
$a,b,\ldots$, respectively. Underlining denotes a tangent 
space
index, {\it i.e.} $A=(\underline{0},\underline{1},
\underline{2},\underline{3})$ whereas $M=(0,1,2,3)$.}
\begin{equation}
\label{eq:Rmatter}
{\cal S}=\frac{1}{{4\kappa}^2}\int d^4x\, \sqrt{-G}
   \left( {}^{(4)}R+{\cal L}_{\cal M} 
   \right) 
\end{equation}
represents $4D$ gravity coupled to matter. 
Here $G_{MN}$, $M,N=0,1,2,3$,
is the $4D$ metric and ${\cal L}_{\cal M}$ is 
the matter Lagrangian. With
an appropriate ansatz (see below), the solutions to  the field 
equations
are then equivalent to those  resulting from an 
action representing $3D$ gravity
coupled to a nonlinear $\sigma$-model
\begin{equation}
\label{eq:Rsigma}
{\cal S}=\frac{1}{{4\kappa}^2}\int d^3x\, \sqrt{|g|}
   \left( {}^{(3)} R-{\cal G}_{\alpha\beta}(\phi^\alpha)
    \partial_{\mu}\phi^\alpha\partial^{\mu}\phi^\beta
   \right), 
\end{equation} 
where the scalar fields $\phi^\alpha$ represent both matter
and $4D$ gravitational fields and 
${\cal G}_{\alpha\beta}(\phi^\alpha)$ 
is the target space metric. In particular this 
equivalence holds for
pure $4D$ gravity if we consider  space-times 
admitting a non-null 
Killing vector $K$. Choosing adapted coordinates 
such that
the corresponding isometry is just a translation, 
{\it i.e.}
$K=K^M\partial_M={\partial}/{\partial y}$,  
the metric may be written in the form
\begin{eqnarray}
\label{eq:metric}
ds^2=G_{MN}dx^Mdx^N=f^{-2}g_{\mu\nu}dx^{\mu}dx^{\nu}+
(-1)^s f^2(dy-\omega_{\mu}dx^{\mu})^2, \ \ f>0,
\end{eqnarray}
where $s=0(1)$ for a space-like(time-like) $K$ and
the components of the $4D$ metric $f,\omega_{\mu}$ and 
$g_{\mu\nu}$ 
are independent of 
the ``fourth coordinate'' $y$. Here $g_{\mu\nu}$ is a
metric on the $3D$ hypersurface coordinatized by $\{x^{\mu}\}$.
This metric will have a Lorentzian(Euclidean) signature 
for reduction
along a space-like(time-like) direction. 
The vector field $\omega_{\mu}$ is defined up to a
gauge transformation $y'=y+\lambda(x^{\mu})$ under which
${\omega'}_{\mu}=\omega_{\mu}-\partial_{\mu}\lambda$. We 
can also
include matter couplings 
consisting of massless scalar and abelian vector fields
if they are independent of  $y$ \cite{Breit}.

The action (\ref{eq:Rsigma}) may be obtained from 
(\ref{eq:Rmatter}) via     
 na\"{\i}ve dimensional reduction followed by dualisation of the 
twist vector $\omega_\mu$ and the vector fields in 
${\cal L}_{\cal M}$.
This action  is invariant under isometries
${\phi^\alpha}'={\phi^\alpha}'(\phi^\beta)$ of 
the $\sigma$-model target space:
Infinitesimally
\begin{equation}
{\phi^\alpha}'=\phi^\alpha+ \xi^\alpha(\phi), 
\end{equation}
 where $ \xi_\alpha $ is a  Killing vector. The full set 
of Killing vectors generate 
the isometries. 
They are  found by solving the  Killing equation
\begin{equation}
\label{eq:Killingeq}
 \xi_{(\alpha:\beta)}=0 
\end{equation}
where $:$ denotes a target space covariant derivative.
The invariance implies that  under an isometry 
transformation a set of solutions $(\phi^\alpha,g_{\mu\nu})$
transforms into a new set of solutions  
$({\phi^\alpha}',g_{\mu\nu})$. 
Starting from one (``seed'') solution an isometry thus 
generates  
a new solution, which  is in general
physically different from the old one ({\it e.g.} a 
stationary
solution can be obtained from a static one).

In the supersymmetric case we expect the above results 
to extend to say that  
$4D$ supergravity/matter systems for a  
certain set of solutions are equivalent 
a $3D$ supersymmetric nonlinear $\sigma$-model
coupled to supergravity. 
We will explicitly demonstrate this below for $N=1,\;4D$ supergravity
by dimensional reduction.
For the supersymmetric $3D$ 
action to be invariant under the
above isometries the
fermionic partners $\chi^\alpha$ of $\phi^\alpha$ must also 
transform under the isometry transformations:
Infinitesimally,
\begin{eqnarray}
\delta\chi^\alpha = \partial_\beta \xi^\alpha \chi^\beta. 
\end{eqnarray}
In particular, a purely bosonic configuration  cannot 
transform into a configuration with fermions under
an isometry. 

In order for a bosonic configuration of supergravity/matter 
to be supersymmetric, {\it i.e.} preserve some supersymmetry, 
there must exist a  spinor $\epsilon$ satisfying
the Killing spinor equations,
\begin{eqnarray} 
\delta_{\epsilon} \Psi_M=0,\;\;\;\;\delta_{\epsilon} \Pi =0,
\end{eqnarray}
where $\Psi_M$ is the Rarita-Schwinger fermion and $\Pi$ 
represents
fermionic matter fields. If the
Killing spinor of the $4D$ configuration  depends on $y$,
the configuration will not be supersymmetric from the $3D$
point of view. We therefore make the extra assumption that 
the $4D$ Killing
spinors are independent of $y$. It is important
to note that even if the bosonic fields are taken to be 
independent
of the isometry direction $y$, a Killing spinor does not 
necessarily 
have this property (examples will be given below).   
The corresponding $3D$ equations obtained by
the dimensional reduction to a supersymmetric $\sigma$-model
are of the form (see {\it e.g.} \cite{Wit93})
\begin{eqnarray}
\label{eq:Killingeq3D}
\delta_{\epsilon} \psi_{\mu}&=& \hat{\cal D}_{\mu}\epsilon=0, \\
\delta_{\epsilon} \chi^{\alpha} &=& 
\partial{\negthickspace\negthickspace / \,}
\phi^\alpha \epsilon=0 , 
\end{eqnarray}
where supersymmetry indices have been suppressed, $\psi_\mu$ 
refers to the $3D$ gravitino and  $\hat{\cal D}_{\mu}$ is a 
derivative containting $3D$ spin- and target space 
connections.  The first equation is invariant
under a target space isometry and the second one   
transforms as a vector; thus $\epsilon$ will satisfy 
$\delta_{\epsilon}\psi_{\mu}'=\delta_{\epsilon} \chi_{\mu}'= 0$.
This proves that $\epsilon$ is a Killing spinor of the 
new configuration if it is so for the original configuration.
Hence, given a {\em supersymmetric}
set of solutions $(\phi^\alpha,g_{\mu\nu},\epsilon)$ we 
can generate a
new {\em supersymmetric} set of solutions 
$({\phi^\alpha}',g_{\mu\nu},\epsilon)$ using  isometry 
transformations
${\phi^\alpha}'={\phi^\alpha}'(\phi^\beta)$.

\section{N=1 Supergravity}

Here, following along the lines of Scherk and Schwarz 
\cite{Sch79,Sch80}, we will 
derive the $3D$ 
supersymmetric action from $4D$  by
dimensional reduction.

We start with the $N=1, D=4$ supergravity 
Lagrangian\footnote{The $4D$ Gamma matrices, denoted by $\Gamma^A$,   
obey the Clifford algebra $\{\Gamma^A, \Gamma^B\}=2\eta^{AB}$.
We also define $\Gamma_5\equiv i\Gamma_{\underline 1}
\Gamma_{\underline 2}\Gamma_{\underline 3}\Gamma_{\underline 0}$ 
and $\Gamma^{AB}\equiv\frac{1}{2} [\Gamma^A,\Gamma^B ]$.}
\cite{Fer76,DesZum},
\begin{equation}
\label{eq:LN=1}
{\cal L}\,=\,\frac{1}{4\kappa^2}ER(\Omega) 
-\frac{i}{2}\varepsilon^{MNPQ}\overline{\Psi}_M\Gamma_5
\Gamma_N{\cal D}_P (\Omega)\Psi_Q,
\end{equation}
where $\Psi_{M}$ is the  gravitino (with Dirac conjugate 
$\overline{\Psi}_M\equiv\Psi^{\dagger}_M i\Gamma^{\underline{0}}$),
$E$ is the determinant of
the vierbein field $E_M^A$, and 
$R(\Omega)\,=\,E^{MA}E^{NB}R_{MNAB}(\Omega) $ is the 
Ricci scalar
($E_A^M$ is the inverse vierbein). 
The Riemann tensor and spin connection are given by
\begin{equation}
R_{MN}^{AB}(\Omega)=\partial_{[M}\Omega_{N]}^{AB}
+\Omega_{[M}^{AD}\Omega_{N]D}^B,
\end{equation}
and
\begin{equation} 
\Omega_{MAB}=\Omega^{0}_{MAB}-\frac{\kappa^2}{2}C_{MAB}, 
\end{equation}
respectively, where $\Omega^0_{MAB}$ is the torsionless spin 
connection of ordinary gravity defined by
\begin{equation}
 \Omega^0_{MAB}=\frac{1}{2}E_A\,^N\partial_{[M}E_{N]B}
-\frac{1}{2}E_A\,^RE_B\,^SE_M\,^D\partial_RE_{SD}-(A\leftrightarrow B)
\end{equation}
and 
\begin{equation}
 C_{NAB}= \overline{\Psi}_N\Gamma_B\Psi_{A}
-\overline{\Psi}_N\Gamma_A\Psi_{B}-\overline{\Psi}_A\Gamma_N\Psi_{B}.
\end{equation}
${\cal D}_P$ is the covariant derivative which acts on the spinors as
\begin{equation}
{\cal D}_M\Psi_N=(\partial_M +\frac{1}{4}\Omega_M\,^{AB}
\Gamma_{AB})\Psi_N.
\end{equation} 
This theory is invariant under local 
supersymmetry $(\epsilon)$, 
local Lorentz $(\lambda)$, and general 
coordinate $(\xi)$ transformations:
\begin{eqnarray}
\label{eq:variE}
\delta E_M\,^A &=& \kappa{\overline \epsilon}\Gamma^A\Psi_M
+E_M\,^B\lambda_{BA}+ \xi^N\partial_NE_M\,^A +\partial_M \xi^N E_N\,^A,\\
\delta \Psi_M &=& \frac{1}\kappa{\cal D}_M(\Omega)\epsilon
+\frac{1}{4}\lambda^{AB}\Gamma_{AB}\Psi_M
+\xi^N\partial_N\Psi_M\,+\,\partial_M\xi^N\Psi_N. 
\end{eqnarray}
There are no auxiliary fields included in this formulation. 
Therefore the supersymmetry  algebra closes 
only modulo the fermionic field equations.

\subsection{Dimensional Reduction}

We choose to consider solutions
with a space-like Killing vector ($s=0$ in (\ref{eq:metric})).
Splitting the 4-indices  $ A=(a,\underline{3}),\;\;M=(\mu,3)$ and
writing the vierbein as
\begin{equation}
\label{eq:Egauge}
\begin{split}
 E_{M}\,^{A} &=
  \begin{pmatrix}
    f^{-1}e_{\mu}\,^{a} & -f\omega_{\mu} \\
    0 & f
  \end{pmatrix},
\end{split}
\end{equation}
the bosonic part of the action is reduced to
\begin{equation} 
\label{eq:L3Dbos}
{\cal S}_{bos}=\frac{1}{{4\kappa}^2}\int d^3x\, e \left[ R-2f^{-2}
\partial_{\mu} f\partial^{\mu} f-\frac{1}{4}f^4 B_{\mu\nu}
B^{\mu\nu} \right],
\end{equation}
where we have defined the twist field
$B_{\mu\nu}=\partial_{[\mu}\omega_{\nu]}$.

To reduce the remaining  part of the 
action it is appropiate to decompose the 
four-component Majorana spinors $\Psi$ above 
into two-component spinors which will become
Majorana spinors $\psi$ in three 
dimensions by choosing the following representation 
for the Clifford algebra (see the appendix for our choice of $3D$
gamma matrices) 
\begin{equation}
\label{eq:Gammadecomp}
\Gamma^a=\gamma^a\otimes{\tau_3}, \hspace{1cm} \Gamma^{\underline{3}}
={\bf 1}\otimes{\tau_2}, \hspace{1cm} 
\Gamma^5=-{\bf 1}\otimes{\tau_1},
\end{equation}
where $\tau_{1,2,3}$ are the  Pauli matrices. 
In this basis $\Psi_M$ takes the form
\begin{equation}
\Psi_M= \left(\begin{array}{clcr}
\psi_1 \\
i\psi_2
\end{array}\right)_M.
\end{equation}
After some simplification  we are left with
\begin{equation}
\begin{split}  
{\cal S}_{f}=\int d^3x\,  & \left[\frac{f^{-1}}{2}
\varepsilon^{\mu\nu\rho}
\varepsilon^{ij}\left(\overline{\psi}^i_{\mu}
\gamma_{\nu}{\cal
D}_{\rho}\chi^j+\overline{\chi}^j\gamma_{\mu}
{\cal D}_{\nu}\psi^i_{\rho}
\right) \right . \\ 
& \left. +\frac{f}{2}\varepsilon^{\mu\nu\rho}\left
( -\omega_{\mu}
\overline{\chi}^i{\cal D}_{\nu}\psi^i_{\rho}
+\omega_{\mu}\overline{\psi}^i_{\rho}{\cal D}_{\nu}\chi^i-
\overline{\psi}^i_{\mu}{\cal D}_{\nu}\psi^i_{\rho}\right)
+ef^{-2}\varepsilon^{ij}\overline{\psi}^i_{\mu}{\chi}^j
\partial^{\mu}f 
\right. \\ 
& \left. +\frac{f^{3}}{4}e\varepsilon^{ij}
\overline{\psi}^i_{\mu}{\chi}^j
\omega_{\nu}B^{\nu\mu}
+\frac{f}{4}e\overline{\psi}^i_{\mu}
\gamma_{\nu}\chi^iB^{\nu\mu}
-\frac{f^{3}}{8}e\varepsilon^{ij}
\overline{\psi}^i_{\mu}
{\psi}^j_{\nu}B^{\mu\nu} \right] +
{\cal S}_{\psi^4} ,
\end{split}
\end{equation}
where $\chi_i\equiv(\psi_i)_3$ and $ {\cal S}_{\psi^4} $ 
represents terms quartic in the fermi fields.

The local transformations (\ref{eq:variE}) can in general 
transform the vierbein fields out of the gauge $E_3^a=0$.
To prevent this, we require $\delta E_{3}^a=0$. This  implies 
\begin{equation}
 \lambda^a_{\ \underline{3}}=
-\kappa f^{-1}\overline{\epsilon}^i\gamma^a\chi^i . 
\end{equation}
Furthermore, $ \delta e_\mu ^a$ can be brought to  canonical form, 
\begin{equation}
 \delta e_\mu ^a=\kappa\overline{\widetilde{\epsilon}^i}
\gamma^a\widetilde\psi_\mu^i , 
\end{equation}
by the following redefinitions,
\begin{eqnarray}
\widetilde{\psi}_{\mu}^{i} &=& f^{1/2}(\psi_{\mu}^{i}
+\omega_{\mu}\chi^{i}+f^{-2}\gamma_{\mu}
\varepsilon^{ij}\chi^{j}) ,\\
 \tilde{\chi}^i &=&f^{-3/2}\chi^{i}, \\
 \tilde{\epsilon}^i &=&f^{1/2}\epsilon^{i},
\end{eqnarray}
provided this is accompanied by a compensating 
Lorentz transformation with parameter
\begin{equation}
\lambda^{ba}=\kappa\overline{\epsilon}^if^{-1}
\gamma^{ba}
\varepsilon^{ij}\chi^j.  
\end{equation}
The reduced action then reads (dropping the tildes) 
\begin{equation}
\label{eq:Stemp}
\begin{split}
{\cal S}=\int d^3x\, & \left\{ \frac{1}{{4\kappa}^2}e
\left[ R-2f^{-2}
\partial_{\mu} f\partial^{\mu}f-\frac{1}{4}f^4 
B_{\mu\nu}B^{\mu\nu}
\right] \right. \\ & \left.
-\frac{1}{2}\varepsilon^{\mu\nu\rho}\overline
{\psi}^i_{\mu}
{\cal D}_{\nu}\psi^i_{\rho}-\frac{1}{8}ef^2
\overline{\psi}^i_{\mu}
\psi^j_{\nu}\varepsilon^{ij}B^{\mu\nu}
-e\overline{\chi}^i D{\negthickspace\negthickspace / \,}
{\chi}^i
+\frac{1}{4}f^2\varepsilon^{\mu\nu\rho}\overline{\chi}^i
\gamma_{\rho}
\chi^j\varepsilon^{ij}B_{\mu\nu}
\right. \\ & \left.
-ef^{-1}\overline{\chi}^i\gamma^\mu 
\partial {\negthickspace\negthickspace / \,}f
{\psi}_{\mu}^j\varepsilon^{ij}
+\frac{1}{4}f^2\varepsilon^{\mu\rho\sigma}
\overline{\psi}^i_{\mu}
\chi^i B_{\rho\sigma}-\frac{1}{2}ef^2
\overline{\chi}^i\gamma_{\mu}
\psi^i_{\nu}B^{\mu\nu} \right\} +
{\cal S}_{\psi^4}.
\end{split}
\end{equation}
In terms of the redefined fields the supersymmetry transformations are
\begin{eqnarray}
\delta e_\mu^a &=&{\kappa}{\overline{\epsilon}}^i
\gamma^a{\psi}_\mu^i \\
\delta f &=&{\kappa}f\overline{\epsilon}^i\chi^j
\varepsilon^{ij} \\
\delta \omega_\mu
&=&-f^{-2}{\overline{\epsilon}}^i{\psi}_\mu^j
\varepsilon^{ij}-
2f^{-2}\overline{\epsilon}^i\gamma_\mu\chi^i \\
\delta\psi_\mu^i &=&{\kappa}^{-1}{\cal D}_\mu
\epsilon^i+
\frac{1}{8\kappa}f^2\varepsilon_{\mu\rho\sigma} 
B^{\rho\sigma}\epsilon^j\varepsilon^{ij}+{\cal O}
(\psi^2
\epsilon)\\
\delta\chi^i &=&\frac{1}{8\kappa}f^2 B^{\mu\nu}
\gamma_{\mu\nu}\epsilon^i-
\frac{1}{2\kappa}f^{-1}\partial {\negthickspace\negthickspace / \,} 
f
\epsilon^j\varepsilon^{ij}+{\cal O}(\psi^2\epsilon).
\end{eqnarray}

As a final step, we will dualise the twist vector 
$\omega_{\mu}$ into
a scalar (the twist potential). 
The field strength $B_{\mu\nu}$ can be treated as 
an independent 
field by adding a Lagrange multiplier field 
$\sigma$ to the 
action (\ref{eq:Stemp}):
\begin{equation}
{\cal S}'={\cal S}+\frac{1}{2}\int d^3x\;\;\sigma
\varepsilon^{\mu\nu\rho}
\partial_{\mu}B_{\nu\rho}.  
\end{equation}
This guarantees that $B_{\mu\nu}$ statisfies the 
Bianchi identity.
The resulting field equation for $B_{\mu\nu}$ is
\begin{equation}
\label{eq:twistpot}
B^{\mu\nu}=-f^{-4}e^{-1}\varepsilon^{\mu\nu\rho}
\partial_{\rho}\sigma
+8\kappa^2 f^{-4}\Pi^{\mu\nu}+{\cal O}(\psi^4).
\end{equation}
where $\Pi^{\mu\nu}$ is a shorthand notation for 
the fermionic
bilinears which are contracted with   $B_{\mu\nu}$ 
in the action
(\ref{eq:Stemp}).  
Inserting this into ${\cal S}'$ and also letting
$\varphi=f^2$ we finally obtain the action for 
a $N=2$ (three dimensional) supersymmetric nonlinear 
$\sigma$-model coupled to  supergravity,  
\begin{equation}
\label{eq:Sfinal}
\begin{split}
{\cal S}'=\int d^3x\, & \left\{ \frac{1}{{4\kappa}^2}e
\left[ R-\frac{1}{2\varphi^2}
(\partial_{\mu} \varphi\partial^{\mu}
\varphi+\partial_{\mu}\sigma\partial^{\mu}\sigma)
\right]
-\frac{1}{2}\varepsilon^{\mu\nu\rho}\overline{\psi}^i_{\mu}
({\cal D}_{\nu}\psi^i_{\rho}-\frac{1}{4\varphi}
{\psi}^j_{\nu}\varepsilon^{ij}
{\partial}_{\rho}\sigma) \right. \\ 
& \left. -e\overline{\chi}^i (D{\negthickspace\negthickspace / \,}
{\chi}^i
-\frac{1}{2\varphi}\partial{ \negthickspace\negthickspace / \,}
\sigma
{\chi}^j\varepsilon^{ij})
-\frac{1}{2}e\varphi^{-1}\overline{\chi}^i\gamma^\mu 
\partial {\negthickspace\negthickspace / \,} \varphi{\psi}_{\mu}^j
\varepsilon^{ij}+\frac{1}{2}\varphi^{-1}e\overline{\chi}^i\gamma^\mu
\partial{ \negthickspace\negthickspace / \,}\sigma{\psi}_{\mu}^i
\right\}+{\cal S}_{\psi^4} .
\end{split}
\end{equation}
The  local supersymmetry transformations which leave 
(\ref{eq:Sfinal}) invariant are
\begin{eqnarray}
\label{eq:sigmatransf}
\delta e_\mu^a &=&\kappa {\overline{\epsilon}}^i
\gamma^a{\psi}_\mu^i \\
\delta \varphi &=&2\kappa \varphi\overline{\epsilon}
^i\chi^j\varepsilon^{ij}  \\
\delta \sigma &=& 2 \kappa \varphi\overline{\epsilon}
^i\chi^i  \\
\delta\psi_\mu^i &=&\kappa^{-1}{\cal D}_\mu\epsilon^i
+\frac{1}{4\kappa}\varphi^{-1}\partial_{\mu}\sigma
\epsilon^j\varepsilon^{ij}+{\cal O}(\psi^2\epsilon) \\
\delta\chi^i &=&\frac{1}{4\kappa}\varphi^{-1}
\partial {\negthickspace\negthickspace / \,}\sigma
\epsilon^i
-\frac{1}{4\kappa}\varphi^{-1}\partial 
{\negthickspace\negthickspace / \,}
\varphi\epsilon^j\varepsilon^{ij}+{\cal O}
(\psi^2\epsilon).
\end{eqnarray}
The commutator of two such transformations gives
local supersymmetry, general coordinate, and Lorentz transformations.
Closure of the algebra requires the use of the Fermi-field equations
of motion since we have no auxiliary fields.

From the action (\ref{eq:Sfinal}) we can read off the 
target space metric. It is given by the line element
\begin{equation}
\label{eq:fsmetric}
ds^2={\cal G}_{\alpha\beta}d\phi^\alpha d\phi^\beta
=\frac{1}{2\varphi^2}(d\varphi^2+d\sigma^2) .
\end{equation}
Changing variables   $z\,=\,\sigma+i\varphi$ 
and ${\overline{z}}\,=\,\sigma-i\varphi\;$, it is easily 
seen that this transforms into
\begin{equation}
{\cal G}_{z\overline z}dzd\overline z=
-\frac{2}{(z-\overline z)^2}dzd\overline z\equiv
\partial_z\partial_{\overline z}K(z,\overline z).
\end{equation}
Hence the target space is a  K\"a{}hler 
manifold with  K\"ahler potential   
$ K(z,\overline z)=-2\,\mbox{ln}(z-\overline z) $.
This manifold is recognized as $SL(2,R)/SO(2)$.

We conclude this subsection by some remarks on the 
time-like reduction ($s=1$ in (\ref{eq:metric})). Starting
from the Lagrangian (\ref{eq:LN=1}) reduction of
the bosonic part will again give (\ref{eq:L3Dbos}) but 
with an opposite sign in front of the $B^2$-term.
However, dualising
the twist vector $\omega_{\mu}$ gives an additional
factor $(-1)^s$
from the contraction of two $\varepsilon$-tensors. The
resulting $\sigma$-model will therefore have 
the same target space metric 
(\ref{eq:fsmetric}) as above. 
Also, the $3D$ fermions will now satisfy a Euclidean Clifford algebra.
The main difference is that there are no Majorana spinors
for this signature of the metric. It is, however, possible to
impose a generalized Majorana condition on the spinors
(see {\it e.g.} \cite{van}). This is
indeed the condition that follows from the $4D$ Majorana condition
by using a decomposition analogous to (\ref{eq:Gammadecomp}).

\subsection{Isometries}
Given the metric (\ref{eq:fsmetric}), 
there are three linearly independent solutions to
the Killing equation (\ref{eq:Killingeq}),
\begin{equation}
\begin{array}{ll} 
\vec{\xi}_1 = 2\varphi{\partial}_\varphi+ 2\sigma{\partial}_\sigma,\\
\vec{\xi}_2 = {\partial}_\sigma, \\
\vec{\xi}_3 = -4\varphi\sigma {\partial}_\varphi
+ 2(\varphi^2-{\sigma}^2){\partial}_\sigma.
\end{array}
\end{equation}
These Killing vectors generate a $3$-parameter group of 
target space isometries and satisfy the $SL(2,R)$ algebra:
\bigskip
\begin{center}
\begin{tabular}{|c|c|c|c|}
  \hline
  $[\vec{\xi}_i,\vec{\xi}_j] $ & $\vec{\xi}_1$ & 
     $\vec{\xi}_2$ & $\vec{\xi}_3$ \\ \hline
  $\vec{\xi}_1$ & $0$ & $-2 \vec{\xi}_2$ &  $2\vec{\xi}_3$\\ \hline
  $\vec{\xi}_2$ & $2\vec{\xi}_2$ & $0$ &$-2\vec{\xi}_1$ \\ \hline
  $\vec{\xi}_3$ & $-2\vec{\xi}_3$ & $2\vec{\xi}_1$ & $0$ \\ \hline
\end{tabular}
\end{center}
\bigskip

Exponentiating the generators one can find 
how finite transformations act on the target space fields;
\begin{equation}
{\phi'}^\alpha = e^{{\lambda}_r{\xi}_r}{\phi}^\alpha,
\end{equation}
(no summation over $r$) where ${\lambda}_r$ is the group 
parameter corresponding 
to the generator $ \xi_r$. 
The finite transformations  are,
\begin{equation}
\label{eq:finiteN=1}
\begin{array}{lll} 
{\xi}_1 :&\varphi'=  e^{2{\lambda}_1}\varphi,\;\;\;\;\;  & 
  {\sigma}'=  e^{2{\lambda}_1}\sigma \\
{\xi}_2 :&\varphi'= \varphi,\;\;\;\; & {\sigma}'=  \sigma+{\lambda}_2\\ 
{\xi}_3:&\varphi'=\frac{\varphi}
{\gamma^2(\varphi^2+\sigma^2)+2\gamma \sigma+1},\;\;\;\;
   &\sigma'=\frac{\gamma(\varphi^2+\sigma^2)+\sigma}
{\gamma^2(\varphi^2+\sigma^2)+2\gamma \sigma+1} .
 \end{array}
\end{equation}
The first two correspond to a global rescaling
of the $4D$ metric and  a trivial gauge transformation of 
the twist potential, respectively.
The third one (Ehlers transform)
is somewhat more complicated.
The transformations (\ref{eq:finiteN=1}) are most 
easily found by noting that
the complex upper half-plane $z=\sigma+i\varphi,\, \varphi>0$ 
with the metric
(\ref{eq:fsmetric}) represents  a $2D$  
surface with constant negative curvature. As is well-known, the
$SL(2,R)$-transformations act on the coordinates as 
\begin{equation}
z'=\frac{\alpha z+\beta }{\gamma z+ \delta},\;\;\;
\alpha\delta-\gamma\beta=1,
\end{equation} 
where $\alpha,\beta,\gamma$ and $\delta$ are real parameters.
Choosing different combinations of the parameters 
we find the above transformations of $\varphi$ and $\sigma$. 
The nontrivial Ehlers transform is given
by $\alpha =1,\;\;\beta=0,\;\;\delta=1$.

For the  full  supersymmetric $\sigma-$model action
to be invariant under  
finite isometry  transformations of $\sigma$ and $\varphi$, 
their fermionic partners must transform as
\begin{equation}
  {\chi'}^i=A{\chi}^i+B{\epsilon}^{ij}{\chi}^j , 
\end{equation}
where
\begin{equation}
A=\frac{\gamma^2(-\varphi^2+\sigma^2)+2\gamma \sigma+1}
{\gamma^2(\varphi^2+\sigma^2)+2\gamma \sigma+1},\;\;\;\;\;
B=\frac{2\gamma^2(\varphi\sigma)+2\gamma \varphi}
{\gamma^2(\varphi^2+\sigma^2)+2\gamma \sigma+1} . 
\end{equation}
These transformations are found by demanding that the 
form of the supersymmetry transformations (\ref{eq:sigmatransf})
are unchanged by the isometries (\ref{eq:finiteN=1}).  
\subsection{Examples}

i) Based on the results above we exemplify the solution-generating 
technique on  plane polarized (pp) gravitational waves. 
These space-times are known to be supersymmetric \cite{Tod83,Hull}.
We have to restrict ourselves 
to  solutions that admit at least one space-like Killing 
vector. With an appropriate choice of the twist potential 
and the three dimensional metric elements our initial metric 
(\ref{eq:metric})
can be cast in the pp-wave solution form:
\begin{equation}
\label{eq:ppmetric}
ds^2={|d\zeta|}^2+2d{U}d{V}+2H(\zeta,\bar{\zeta})d{U}^2  
\end{equation}
where, 
\begin{equation}
\zeta=x^1+ix^2,\;\;\;{U}=\frac{1}{\sqrt{2}}(x^3-x^0),\;\;\;{V}=
\frac{1}{\sqrt{2}}(x^3+x^0),  
\end{equation}
and $H(\zeta)$ is  a harmonic function, 
$\partial_{\zeta}\partial_{\bar{\zeta}}H=0$.
From the $4D$ Killing spinor equation,
${\cal D}_M\epsilon =0$,
it can be shown that the Killing spinor is independent of $x^3$ 
(in fact the Killing spinor is constant in this case).
Comparing the metrics (\ref{eq:metric}) and (\ref{eq:ppmetric}) and  
using $\varphi=f^2$ we find the relations
\begin{equation}
H=\frac{\varphi-1}{2},\;\;\;\;\omega_0=1-\varphi^{-1} .  
\end{equation}
All functions depend on $\zeta$ and $\bar{\zeta}$ only.
An isometry transformation 
(see (\ref{eq:finiteN=1}))
results in a new pp-wave solution characterized by the
transformed harmonic function $H'=\frac{1}{2}(\varphi'-1)$. 
Note also that differentiating
the second equation and using (\ref{eq:twistpot})
one finds the Cauchy-Riemann equations; {\it i.e.}
$z=\sigma+i\varphi$ is an analytic function of $\zeta$.
This requirement will solve the $3D$ Killing spinor equation
$\delta_{\epsilon} \chi^i=0$ \cite{How96} (also, the integrability
condition for $\delta_{\epsilon} \psi^i=0$ is satisfied for any
static $3D$ metric $h_{\mu\nu}$).   
\\ \noindent
ii) A simple example which illustrates the importance of
the assumption that the Killing spinors are independent 
of the isometry direction is flat space in polar coordinates,
\begin{equation}
 ds^2=-dt^2+d\rho^2+\rho^2d\theta^2+dx^3dx^3,\;\;\;\;
\rho^2=(x^1)^2+(x^2)^2,\; \mbox{tan}\,\theta =x^2 /x^1. 
\end{equation}
With the choice $\varphi=f^2=\rho^2,\;\omega_\mu=0$ and 
$g_{\mu\nu}=\rho^2 \eta_{\mu\nu}$, 
this is of the form (\ref{eq:metric}) with $\theta$ as  
the isometry direction. The Killing spinor equation is solved
by $\epsilon=e^{\frac{1}{2}\Gamma_{\underline{1}\,\underline{3}}\theta}
\epsilon_0$ where $\epsilon_0$ is a constant spinor.
Performing a nontrivial target space isometry transformation, using the
third equation in (\ref{eq:finiteN=1}), we
generate $\varphi'={\rho^2}/({\gamma^2\rho^4+1})$. The Ricci scalar for 
this solution is nonvanishing. Therefore it can neither be
a  flat space-time nor a pp-wave because these have $R=0$.
Since these are the only possible $N=1$ supersymmetric space-times 
we conclude that the generated solution is not supersymmetric.  

\section{N=2 Supergravity}
In this section $N=2$ supergravity 
subject to the  isometry assumptions given in 
section 2 is considered. 
We anticipate that  $N=2,\;\;D=4$ 
supergravity, when reduced to $3D$, will 
consist of an $N=4$ locally 
supersymmetric nonlinear $\sigma$-model. 
 The signature 
of the $3D$ submanifold will be Lorentzian or Euclidean 
depending on whether the reduction has been performed 
along a time- or space-like direction. To ensure $N=4$
supersymmetry the $\sigma$-model target space is required 
to have quaternionic structure. Our course of action 
will consist of  dimensional reduction of the bosonic part of the 
Lagrangian only. Using  the isometries of 
the target space manifold we can then
generate new supersymmetric solutions.

The   $N=2,\;\;D=4$ supergravity 
Lagrangian is \cite{Fer76b}
\begin{eqnarray}
\label{eq:sugrN2}
{\cal L}&=&\frac{E}{4{\kappa}^2}R-
\frac{E}{2}\overline{\Psi}_M^i\Gamma^{MNP}D_N\Psi_P^i-
\frac{E}{4}F^2_{MN} \nonumber\\
&-&\frac{\kappa}{4 \sqrt{2}}\overline{\Psi}_M^i[(F^{MN}+
{\hat F}^{MN})
+\frac{\Gamma_5}{2}({\tilde{F}}^{MN}+
\tilde{{\hat F}}{}^{MN})]{\Psi}_N^j\epsilon^{ij},
 \end{eqnarray}
where $F_{MN}$, ${\tilde F}_{MN}$ and ${\hat F}_{MN}$ denote the
electromagnetic, dual, and the supercovariantized 
field strength, respectively. Its bosonic part
is the Einstein-Maxwell Lagrangian.   
\subsection{Reduction}
The  dimensional reduction of the bosonic part of the 
Lagrangian (\ref{eq:sugrN2}) is done along the lines of section 3.
Since the most interesting solutions contain a 
time-like Killing vector, we will focus on this 
case. This amounts to having a  reduced theory with a Euclidean  
signature of the $3D$ metric. As remarked in the 
previous section, the pure gravitational
part will give (\ref{eq:L3Dbos}) with a different sign in front of
the $B^2$-term. The assumption that $F_{MN}$ is time-independent
enables us to  introduce  electric and 
magnetic potentials $v$ and $u$ through
\begin{eqnarray}
F_{\mu0}&=&\frac{1}{\sqrt{2}}\partial_\mu v,\\
F_{\mu \nu}&=&\frac{\varphi}{\sqrt{2}}e\varepsilon_
{\mu\nu\rho}\partial^\rho u.
\end{eqnarray}
The twist potential $\sigma$ is now defined by the relation
\begin{equation}
\label{eq:ein}
\partial_\mu \sigma+v\partial_\mu u-u\partial_\mu v=-\varphi^2 
e\varepsilon_{\mu\nu\rho}\partial^\nu \omega^\rho,
\end{equation}
to ensure consistency with the Einstein equations \cite{Isr72}.

The reduction of the bosonic part of (\ref{eq:sugrN2}) can then be 
cast into the form  (\ref{eq:Rsigma}) where the   
$\sigma$-model target space metric is given by
\begin{equation}
\label{eq:sigmaN2}
ds^2 =\frac{1}{2\varphi^2}\left\{ d\varphi^2+(d\sigma+vdu-udv)^2\right\}
-\frac{1}{\varphi}\left\{dv^2+du^2\right\}.
\end{equation} 
This manifold is an Einstein space with constant negative
curvature; the Ricci tensor satisfies:
\begin{equation}
  {\cal R}_{\alpha\beta}=-\frac{1}{4}(8+d){\cal G}_{\alpha\beta},
\end{equation}
where $d=4k$ is the number of scalars (four in our case).
In fact, this noncompact manifold can be identified 
as $SU(2,1)/S(U(1,1)\times U(1))$ \cite{Breit}.
Its holonomy group is contained in $Sp(k)\times Sp(1)$ \cite{BagWitten83}.
Manifolds that have these properties are called quaternionic.
This is precisely the structure needed   on the target space in order
to have an $N=4,\;D=3$ supersymmetric nonlinear $\sigma$-model coupled to 
supergravity \cite{Wit93}.

\subsection{Isometries}
We now turn to  the isometries of the $\sigma$-model.
Solving the Killing vector equation (\ref{eq:Killingeq}) we find
that the target space  (\ref{eq:sigmaN2}) possesses  
  $ 8 $ Killing vectors :
\begin{align}
\vec{\xi}_1 &= 2\varphi{\partial}_\varphi+ 2\sigma{\partial}_\sigma
+ u{\partial}_u+ v{\partial}_v,\\
\vec{\xi}_2 &= v{\partial}_\sigma+ {\partial}_u, \\
\vec{\xi}_3 &=  u{\partial}_\sigma- {\partial}_v,\\
\vec{\xi}_4 &=  {\partial}_\sigma,\\
\vec{\xi}_5 &=  v{\partial}_u-u {\partial}_v,\\
\vec{\xi}_6 &=  2\varphi u{\partial}_\varphi+[u\sigma
+\frac{v}{2}(u^2+v^2)-v\varphi]\partial_\sigma
+\frac{1}{2}(u^2-3v^2+2\varphi)\partial_u+(2uv-\sigma)\partial_v,\\
\vec{\xi}_7 &=  2\varphi v{\partial}_\varphi+[v\sigma
-\frac{u}{2}(u^2+v^2)+u\varphi]\partial_\sigma+(2uv
+\sigma)\partial_u+\frac{1}{2}(v^2-3u^2+2\varphi)\partial_v,\\
\vec{\xi}_8 &=  -4\varphi\sigma{\partial}_\varphi
+[-2\varphi(u^2+v^2)+\frac{1}{2}(u^2+v^2)^2
+2(\varphi^2-\sigma^2)]\partial_\sigma\nonumber\\
&-(2u\sigma+u^2v+v^3-2\varphi v)\partial_u
+(-2v\sigma+uv^2+u^3-2\varphi u)\partial_v,
\end{align}
which satisfy the $SU(2,1)$ isometry algebra:
\bigskip 
\begin{center}
\begin{tabular}{|c|c|c|c|c|c|c|c|c|}
  \hline
  $[\vec{\xi}_i,\vec{\xi}_j] $ & $\vec{\xi}_1$ & $\vec{\xi}_2$ 
& $\vec{\xi}_3$ & $\vec{\xi}_4$ & $\vec{\xi}_5$ & $\vec{\xi}_6$ 
& $\vec{\xi}_7$ & $\vec{\xi}_8$\\ \hline
  $\vec{\xi}_1$ & $0$ & $-\vec{\xi}_2$ &  $-\vec{\xi}_3$ 
& $-2\vec{\xi}_4$ & $0$ & $\vec{\xi}_6$ & $\vec{\xi}_7$ 
& $2\vec{\xi}_8$\\ \hline
  $\vec{\xi}_2$ & $ $ & $0$ & $2\vec{\xi}_4$ & $0$ 
&$\vec{\xi}_3$ & $\vec{\xi}_1$ & $3\vec{\xi}_5$ 
& $-2\vec{\xi}_7$ \\ \hline
  $\vec{\xi}_3$ & $ $ & $ $  & $0$ & $0$ & $-\vec{\xi}_2$ 
& $3\vec{\xi}_5$ & $-\vec{\xi}_1$  & $-2\vec{\xi}_6$ \\ \hline
 $\vec{\xi}_4$ & $ $ & $ $ &$ $ & $0$ & $0$ & $\vec{\xi}_3$
& $\vec{\xi}_2$ & $-2\vec{\xi}_1$   \\ \hline
 $\vec{\xi}_5$ & $ $ & $ $ & $ $ &$ $ & $0$ & $\vec{\xi}_7$ 
& $-\vec{\xi}_6$ & $0$ \\ \hline
 $\vec{\xi}_6$ & $ $ & $ $ & $ $ &$ $ &$ $&  $0$ & $\vec{\xi}_8$ 
& $0$ \\ \hline
 $\vec{\xi}_7$ & $ $ & $ $ & $ $ &$ $ &$ $& $ $& $0$ & $0$ \\ \hline
 $\vec{\xi}_8$ & $ $ & $ $ & $ $ &$ $ &$ $& $ $& $ $ & $0$ \\ \hline
\end{tabular}
\end{center}
\bigskip

It is convenient to introduce Ernst potentials ${\cal E}=\Xi +i\sigma$ and 
$\Phi=\frac{1}{\sqrt{2}}(u+iv)$ where $\Xi=\varphi-\frac{1}{2}(u^2+v^2)$.
Then the finite transformations are given by \cite{Kra69,Kin73}:
\begin{equation}
\label{eq:finiteN=2}
\begin{array}{rll}
I. \;\;\;&{\cal E}'=\alpha \bar{\alpha}{\cal E},&\Phi'
=\alpha \Phi\\
II. \;\;\;&{\cal E}'={\cal E}+ib,&\Phi'=\Phi\\
III.\;\;\;&{\cal E}'={{\cal E}}/({1+ic{\cal E}}),&\Phi'
={\Phi}/({1+ic{\cal E}})\\
IV. \;\;\;&{\cal E}'={\cal E}-2\bar{\beta}\Phi
-\bar{\beta}\beta,&\Phi'=\Phi+\beta\\
V. \;\;\;&{\cal E}'={{\cal E}}/({1-2\bar{\gamma}\Phi
-\gamma\bar{\gamma}{\cal E}}),&\Phi'=({\Phi
+\gamma{\cal E}})/({1-2\bar{\gamma}\Phi-\gamma\bar{\gamma}{\cal E}})
\end{array}
\end{equation}
The parameters $b$, $c$ are real and $\alpha,\;
\beta,\;\gamma$ are complex.

The first transformation  is  
an electromagnetic duality rotation for $|\alpha|=1$. 
If the modulus of $\alpha$ is different from unity, the effect of
this transformation may be shown to be a scaling, 
$ds^2\rightarrow  |\alpha|^2 ds^2$, of the $4D$ metric.
 The second 
and fourth are gauge transformations of the potentials. 
Only the third and  fifth transformations are nontrivial. 
\subsection{Examples}

i) In our first example we use static supersymmetric space-times 
as ``seed'' solutions. It is well known that these 
must be of the Papapetrou-Majumdar type \cite{Tod83,Gib82},
\begin{equation}
\label{eq:papa}
 ds^2=-f^2dt^2+f^{-2}\delta_{\mu\nu}dx^\mu dx^\nu,\;\;\;\;f
=\frac{v}{\sqrt{2}}. \end{equation}
In accordance with equation  (\ref{eq:ein}) we put $u=\sigma=0$.
The $N=2,\;4D$ Killing spinor equation is
\begin{equation}
\label{eq:KillingN=2}
\partial_M\epsilon+\frac{1}{4}\Omega_{M}^{0\ AB}\Gamma_{AB}\epsilon
    +\frac{1}{4}F_{NP}\Gamma^{NP}\Gamma_M\epsilon=0,
\end{equation}
where $\epsilon$ is a Dirac spinor. This equation has a  
solution $\epsilon$, subject to the constraint 
$\Gamma^{\underline{0}}\epsilon=\epsilon$, 
such that $\partial_t\epsilon=0$.
Supersymmetry will then be preserved under the transformations I-V.

The  finite transformations III leave the solution invariant. 
The transformations V give:
\begin{eqnarray}
u'&=&\frac{-\sqrt{2}\gamma_1 v^2}{(1-\sqrt{2}\gamma_2 v)^2
+2(\gamma_1 v)^2},\;\;\;
\;\;\;v'=\frac{v(1-\sqrt{2}\gamma_2 v)}{(1-\sqrt{2}\gamma_2 v)^2
+2(\gamma_1 v)^2},\nonumber\\
\varphi'&=&\frac{2\gamma_1^2 v^4+v^2(1-\sqrt{2}\gamma_2 v)^2}
{[(1-\sqrt{2}\gamma_2 v)^2+2(\gamma_1 v)^2]^2},\;\;\;
\gamma=\gamma_1+i\gamma_2.
\end{eqnarray}
For $\gamma_1=0$ we generate a solution of the type (\ref{eq:papa}). 
These transformation belongs to the ``static'' subgroup $SL(2,R)$ 
of the full isometry group $SU(2,1)$.
On the other hand starting from an electrically charged 
solution, a general $\gamma$ generates a nonzero magnetic potential. 
From equation
(\ref{eq:ein}), we see that the new solution is not necessarily 
static. Rather it belongs to the more general 
class of stationary supersymmetric solutions known 
as Israel-Wilson-Perjes (IWP) metrics \cite{Isr72}. 
Note that if the original 
solution is asymptotically flat so will  the  new one be.
\\ \noindent
ii) Next we start from the Robinson-Bertotti solution \cite{Exact}. 
The Robinson-Bertotti solution is the only metric
belonging to the IWP class   which 
preserve all supersymmetries. 
 The metric can be written in the form
\begin{equation}
ds^2=(1-\lambda y^2)dx^2+(1-\lambda y^2)^{-1}dy^2+
(1-\lambda z^2)^{-1}dz^2-(1+\lambda z^2)dt^2.
\end{equation}    
The Maxwell  field strength is constant and 
by an electric-magnetic  duality transformation  
can be taken to be purely electric,
\begin{equation}
F_{30}=\sqrt{2\lambda}, \;\;\lambda={\rm constant}.
\end{equation}
A crucial property of this class of solutions is that the Weyl 
tensor vanishes.

Choosing $\varphi=1+\lambda z^2$ and  
$g_{\mu\nu}=diag(\varphi (1-\lambda y^2),\varphi (1-\lambda y^2)^{-1},1)$, 
we get the Robinson-Bertotti metric  in the form (\ref{eq:metric}). 
Applying the transformation III the fields transform into
\begin{eqnarray}
u'&=&\frac{2c\sqrt{\lambda} z \Xi}{1+(c \Xi)^2},\;\;\;
v'=\frac{2\sqrt{\lambda} z}{1+(c \Xi)^2},\nonumber\\
\varphi'&=&\frac{\varphi}{1+(c \Xi)^2},\;\;\;\sigma'
=\frac{-c \Xi^2}{1+(c \Xi)^2},\;\;\;\Xi=1-\lambda z^2.
\end{eqnarray}
As can be seen, we have generated a magnetic field as well as a rotation.
Since this  solution has a nonvanishing Weyl tensor, it 
does not belong to the Robinson-Bertotti class of solutions.
We thus conclude that some supersymmetry is
broken. However, examining the Killing spinor equation
(\ref{eq:finiteN=2})
one finds that there are no solutions satisfying 
$\partial_t\epsilon=0$; which is a crucial  argument 
required to guarantee the supersymmetry of the generated solution. 
This is in agreement with our general discussion in section 2.
\\ \noindent
iii) Another possibility is to start with a pp-wave as ``seed'' 
solution. The supersymmetry of these solutions 
has been discussed by Hull  \cite{Hull}.
With a metric of the form (\ref{eq:ppmetric}) this solution is 
supersymmetric provided that $ F_{MN} $ vanishes. Furthermore 
the Killing spinor is time-independent (in fact constant). 
The transformations I-III are exactly those considered in 
the first example in section 3.3 and apart from the
trivial transformation IV the only new feature 
comes from the last transformation, V. It generates a nonvanishing
field strength, implying that the new solution is not a pp-wave.

\section{Discussion}
In this article we have shown how to generate supersymmetric 
solutions to  $N=1$ and $N=2,\;\;4D$ supergravities using 
known techniques employed in ordinary general relativity. After 
a  na\"{\i}ve  dimensional reduction of $N=1,\;4D$ supergravity 
we have explicitly 
obtained $N=2,\;3D$ supergravity plus a nonlinear supersymmetric
$\sigma$-model. For $N=2$ we have reduced the bosonic
part only. Since the target space metric of the resulting 
$\sigma$-model 
has a quaternionic structure it can be extended
to an $N=4$ nonlinear supersymmetric $\sigma$-model.
Examining the Killing spinor equation we have shown that the
supersymmetries of the solutions are preserved under the target
space isometry transformations, provided that the Killing spinor 
is independent of the coordinate along which the reduction is
done.        

Supersymmetric solutions of supergravity 
theories are usually discussed in the context of
consistently truncated Lagrangians. It means that some bosonic 
fields are set to zero in the 
full supersymmetric Lagrangian
such that the solutions
of the truncated systems are also solutions, albeit specific, 
of the untruncated theory \cite{stel}.
The $\sigma$-model resulting from the reduction of the
truncated supergravity Lagrangian is  a consistent truncation of 
the full supersymmetric nonlinear $\sigma$-model obtained from
the reduction of the untruncated $4D$ supergravity. Since the target space
 ${\cal T}$ of a consistently truncated $\sigma$-model  
is a submanifold
of the target space  ${\cal T}'$ of the untruncated supersymmetric 
$\sigma$-model, the isometries of   ${\cal T}$ will also be isometries 
of ${\cal T}'$ \cite{Breit}.
Our method would therefore be applicable  
for all consistently truncated $4D$ supergravities. 

The solution-generating method used in \cite{Ehl57}-\cite{Jen95}
applies also to higher 
dimensional systems consisting of 
an antisymmetric tensor field coupled to 
$D$-dimensional dilaton-gravity. 
Various  solutions such as black $p$-branes  
have been generated in this way \cite{Gal98}.
Our investigations suggest that the solution-generating 
technique in such systems can also be extended along the
lines described in this paper to include supersymmetry. 
This we will discuss elsewhere \cite{prog}.

\vglue 0.6cm
{\bf \noindent  Acknowledgements \hfil}\\
We are grateful to Ulf Lindstr\"om and Homayoun Hamidian 
for comments and reading the manuscript. 
We also thank Chris Hull, Bernard de Wit
and Ingemar Bengtsson for  useful comments.
\vglue 0.4cm

\renewcommand{\thesection}{Appendix:}
\setcounter{section}{0}
\renewcommand{\theequation}{\Alph{section}.\arabic{equation}}
\setcounter{equation}{0}

\section{3D Gamma Matrices}
The $2+1$ dimensional 
gamma matrices $\gamma^a$ satisfy
\begin{equation} 
\label{eq:3DCliff}
\{\gamma^a, \gamma^b\}=2\eta^{ab}, \hspace{1cm}
\eta^{ab}=\mbox{diag}(-1,1,1).
\end{equation}
In $2+1$ dimensions there are two inequivalent irreducible 
representations of the Clifford algebra 
(\ref{eq:3DCliff}) and we choose the one for which 
$\gamma^{[a}\gamma^b\gamma^{c]}=
\varepsilon^{abc}$. A specific (real) representation in terms of the
Pauli matrices $\tau_{1,2,3}$ is $\gamma^{\underline{0}}=i\tau_2,
\gamma^{\underline{1}}=\tau_1,\gamma^{\underline{2}}=\tau_3$.
We also define 
\begin{equation}
\gamma^{ab}=\frac{1}{2}[\gamma^a,\gamma^b ]=\varepsilon^{abc}\gamma_c.
\end{equation}
The Dirac conjugate of a spinor $\psi$ is 
$\overline{\psi}\equiv\psi^{\dagger} i\gamma^{\underline{0}}$. 
A Majorana spinor $\psi$ obeys the reality condition
$\overline{\psi}=\psi^c$, where $\psi^c\equiv\psi^T C$
is the Majorana conjugate 
and  $C$ the charge conjugation matrix  defined  through 
\begin{equation}
C=-C^{-1}, \hspace{2cm} C\gamma^a C^{-1} =-(\gamma^a)^T.
\end{equation} 
From these properties, the ``flip'' identities
\begin{equation}
\overline{\psi}\gamma^{a_1}\ldots\gamma^{a_k} \chi = (-1)^k
\overline{\chi} \gamma^{a_k}\ldots\gamma^{a_1}\psi, 
\end{equation}
follow for any two Majorana spinors $\psi$ and $\chi$.
It is convenient to choose $C=\tau_2$.
Bilinears of the type 
 $\overline{\psi}\chi$ and $\overline{\psi}\gamma^a\chi$
will then be real.

\newpage

\eject
\setlength{\baselineskip}{15pt}

\end{document}